\documentclass[preprint,showpacs,showemail,preprintnumbers,pre]{revtex4}
\usepackage{graphicx}
\usepackage{dcolumn}
\usepackage{bm}
\usepackage{amssymb,amsmath}

\begin{document}
\title{
Dynamic phase transition in the two-dimensional kinetic Ising model in an 
oscillating field: Universality with respect to the stochastic dynamic
}

\author{
G.~M.~Buend{\'{\i}}a$^{1,2}$}\email{buendia@usb.ve}
\author{
P.~A.~ Rikvold$^{3,4}$}\email{prikvold@fsu.edu}

\affiliation{
$^1$Department of Physics, Universidad Sim{\'o}n Bol{\'{\i}}var, Caracas 1080, 
Venezuela\\
$^2$ Consortium of the Americas for Interdisciplinary Science, University of 
New Mexico, Albuquerque, New Mexico 87131, USA\\
$^3$Center for Materials Research and Technology
and Department of Physics,
Florida State University, Tallahassee, Florida 32306-4350, USA\\
$^4$ National High Magnetic Field Laboratory, Tallahassee, Florida 32310, USA
}
\date{\today}

\begin{abstract}
We study the dynamical response of a two-dimensional Ising model subject to a 
square-wave oscillating external field. In contrast to earlier studies, 
the system evolves under a so-called 
soft Glauber dynamic [P.~A.\ Rikvold and M.~Kolesik, J.\ Phys.\ A: Math.\
Gen.\ {\bf 35}, L117 (2002)], for which both nucleation and interface
propagation are slower and the interfaces smoother
than for the standard Glauber dynamic. 
We choose the temperature and magnitude of the external 
field such that the metastable decay of the system following field
reversal occurs through nucleation 
and growth of many droplets of the stable phase, i.e., the  
multidroplet regime.
Using kinetic Monte Carlo simulations, 
we find that the system undergoes a nonequilibrium phase transition, in 
which the symmetry-broken dynamic
phase corresponds to an asymmetric stationary limit 
cycle for the time-dependent magnetization. The critical point  
is located where the half-period of the external field is approximately 
equal to the metastable lifetime of the system. We employ  
finite-size scaling analysis to investigate the 
characteristics of this dynamical phase transition.  
The critical exponents and the fixed-point value of the 
fourth-order cumulant are found to be 
consistent with the universality class of the two-dimensional equilibrium 
Ising model. As this universality class has previously been established
for the same nonequilibrium model evolving under the standard Glauber dynamic, 
our results indicate that this far-from-equilibrium phase 
transition is universal with respect to the choice of the stochastic dynamics.

\end{abstract}


\maketitle


\section{Introduction}
\label{sec:INTRO}

Kinetic Ising or lattice-gas 
models with stochastic dynamics have been successfully applied 
to study a number of dynamical physical phenomena, including metastable decay 
\cite{RIKV94A,RAMO99,BERT04,BERT04B,FRAN05,FRAN06}, 
hysteretic responses \cite{SIDE98,SIDE99,KORN00}, 
and magnetization switching in nanoscale ferromagnets
\cite{RICH94,NOVO02A}. 
Among the dynamic phenomena in such models that have attracted particular
attention in recent years, is the dynamic phase transition (DPT) observed in 
systems with Ising-like symmetry that are driven 
far from equilibrium by an oscillatory applied 
force (typically a magnetic field or (electro)chemical potential).
This phenomenon was first 
observed in kinetic simulations of a mean-field model \cite{TOME90,MEND91} 
and later studied intensively by 
mean-field \cite{ZIMM93A,BUEN98,ACHA95,CHAK99}, 
Monte Carlo \cite{LO90,ACHA95,CHAK99,SIDE98,SIDE99,KORN00,KORN02B,ROBB07}, 
and analytical \cite{FUJI01,TUTU04,MEIL04,DUTT04} methods. 
In this transition, the dynamic order parameter, which is the cycle-averaged 
magnetization, vanishes in a singular fashion at a critical value of the 
period of the applied field. 
Recently, strong experimental evidence has emerged that this nonequilibrium 
phase transition is observable in magnetic thin-film systems \cite{ROBB08}, 
and an analogous phenomenon has been observed in simulations of 
a model of the heterogeneous catalytic oxidation of CO \cite{MACH05,BUEN06C}. 

Perhaps the most fascinating aspect of this far-from-equilibrium phase 
transition is that it belongs to the same universality class as the 
corresponding equilibrium Ising model. This result is predicted from 
symmetry arguments \cite{GRIN85,BASS94} and has been confirmed by 
exhaustive kinetic Monte Carlo simulations 
\cite{SIDE98,SIDE99,KORN00,KORN02B} and analytical results \cite{FUJI01}. 
Very recently, the field conjugate to the dynamic order parameter was 
identified as the cycle-averaged applied field, 
and a fluctuation-dissipation relation valid near the 
nonequilibrium critical point was numerically established \cite{ROBB07}. 

The physics of equilibrium phase transitions 
is well understood, and it is well established that structures arising from 
different dynamics that obey detailed balance and respect the same 
conservation laws exhibit universal asymptotic large-scale features. However, 
the mechanisms behind nonequilibrium phase transitions are not that well known, 
and the dependence on the specific dynamic is still an open question. 
Except for the study of the model of CO oxidation \cite{MACH05,BUEN06C}, 
all previous kinetic Monte Carlo simulations in which this DPT 
was observed, were performed with the standard 
stochastic Glauber \cite{GLAU63} or Metropolis \cite{METR53} dynamics. 
All these studies,
including the study of CO oxidation which used a very different dynamic,  
found critical exponent ratios consistent with the equilibrium Ising values, 
$\gamma/\nu = 7/4$ and $\beta/\nu = 1/8$. This gives a reasonable indication 
that this DPT is universal with respect to details of the model 
and the stochastic dynamics. 
However, a more direct test of just the universality with respect to the 
dynamics would be to use a significantly different stochastic dynamic for the 
two-dimensional kinetic Ising model. 
Such a test is the subject of the present paper. 

All stochastic dynamics that respect detailed 
balance eventually lead to thermodynamic equilibrium \cite{LAND00}, 
and all dynamics that obey the same conservation laws also give the 
same long-time dynamics (e.g., a $t^{1/2}$ dependence of the characteristic 
length for phase ordering with a non-conserved order parameter and 
a $t^{1/3}$ dependence for phase separation with a conserved order parameter 
\cite{GUNT83B}). 
However, it has recently been demonstrated that different stochastic dynamics 
give quantitatively dramatically different results for low-temperature 
nucleation \cite{PARK04,BUEN04B}, 
as well as for the nanostructure and mobility of field-driven interfaces 
\cite{RIKV02,RIKV02B,RIKV03,BUEN06,BUEN06B}. 
The differences are particularly striking between dynamics known as 
``hard," in which the effects of the configurational 
and field-related (``Zeeman") energy contributions in the transition 
rate do not factorize, and ``soft," for which such factorization is possible 
\cite{RIKV02,PARK04,MARR99}. The class of hard dynamics includes the 
standard Glauber and Metropolis dynamics, while the soft dynamics here will 
be represented by the ``soft Glauber dynamic" introduced in 
Ref.~\cite{RIKV02}, whose transition rate is given in Sec.~\ref{sec:MODEL}. 
Briefly, the nanostructure of field-driven ``hard" interfaces is 
characterized by a local interface width and mobility that increase 
dramatically with the strength of the applied field, 
while ``soft" interfaces remain relatively smooth and slow-moving, independent 
of the field \cite{RIKV02}. Similarly, low-temperature nucleation under 
hard dynamics becomes very fast for strong fields, while under 
soft dynamics it remains thermally activated and thus very slow, even for 
very strong fields \cite{PARK04}.  
While the soft Glauber dynamic is probably not particularly 
relevant to any specific 
physical system, it is ideally suited for comparison with the standard, 
hard Glauber dynamic in investigating universal properties of the DPT. 

The rest of this paper is organized as follows. 
The kinetic  Ising model and its dynamics are introduced in 
Sec.~\ref{sec:MODEL},
and the numerical results and finite-size scaling analysis are 
presented in in Sec.~\ref{sec:MC}. 
A summary and conclusions are given in Sec.~\ref{sec:DISC}.

\section{Model and Dynamics}
\label{sec:MODEL}
For this study we choose a kinetic, nearest-neighbor, Ising ferromagnet on a 
square lattice with
periodic boundary conditions. The Hamiltonian is given by
\begin{equation}
 {\cal H}= -J\sum_{\langle ij \rangle} s_i s_j - H(t)\sum_{i} s_{i},
\label{eq:Ham}
\end{equation}
where $s_{i}=\pm1$ is the state of the spin at the site $i$, $J>0$ is the 
ferromagnetic interaction, $\sum_{\langle ij \rangle}$ 
runs over all nearest-neighbor 
pairs, $\sum_{i}$ runs over all $L^2$ lattice sites, and $H(t)$ is an 
oscillating, spatially uniform applied field. The magnetization per site 
\begin{equation}
 m(t)=\frac{1}{L^2}\sum_{i=1}^{L^2}s_{i}(t),
\label{eq:mag}
\end{equation}
is the density conjugate to $H(t)$. The temperature $T$ 
(in this paper given in units such that Boltzmann's constant equals
unity) is fixed below its 
critical value ($k_{B}T_{c}= J/ \ln (1+\sqrt{2})$), 
so that, when there is no external field,
the system has two degenerate equilibrium states with magnetizations of equal 
magnitude and 
opposite direction. When an external field is applied the degeneracy is 
lifted, and the equilibrium state is the one with magnetization in the same 
direction as the field. If the external field is not too strong, the state with 
opposite magnetization direction is metastable and eventually decays toward 
equilibrium \cite{RIKV94A}. 
This model is equivalent to a lattice-gas model with local occupation
variables $c_i = (s_i+1)/2 \in \{0,1\}$ and (electro)chemical potential 
$\mu \propto H$ (for further details, see Ref.~\cite{RIKV02B}).

The system evolves under the soft Glauber single-spin-flip 
(non-conservative) stochastic dynamic with updates at randomly chosen sites. 
In the lattice-gas representation, this corresponds to
adsorption/desorption without lateral diffusion.  The time unit is 
one Monte Carlo step per spin (MCSS). When the system is in contact with a 
heat bath at a temperature $T$, each proposed spin flip 
is accepted with probability
\begin{equation}
 W_{\rm SG}
= \frac{1}{1+ \exp (\beta\Delta E_J)}\frac{1}{1+ \exp (\beta\Delta E_H)}\;.
\label{eq:softg}
\end{equation}
Here $\beta = 1/T$, $\Delta E_J$ is the energy change corresponding to the 
interaction
term, and $\Delta E_H$ is the energy change corresponding to the field 
term in the Hamiltonian, Eq.~(\ref{eq:Ham}).
This transition probability 
is to be contrasted with those of the standard, hard Glauber dynamic, 
\begin{equation}
 W_{\rm HG}
= \frac{1}{1+ \exp (\beta\Delta E)}\;,
\label{eq:hardg}
\end{equation}
and the Metropolis dynamic,
\begin{equation}
 W_{\rm M}
= {\rm Min} [1, \exp (-\beta\Delta E)]\;,
\label{eq:metro}
\end{equation}
where $\Delta E = \Delta E_J + \Delta E_H$ is the total energy change that 
would result from a transition. 

The dynamical order parameter is the time-averaged magnetization over 
the $k$th cycle of the oscillating field \cite{TOME90},
\begin{equation}
Q_k =\frac{1}{2t_{1/2}} \int_{(k-1)2t_{1/2}}^{k2t_{1/2}} dt \, m(t) \;,
 \label{eq:Q}
\end{equation}
where $t_{1/2}$ is the half-period of the applied
field. The cycle is chosen such that it starts when $H(t)$ changes sign. 
We also measured the normalized cycle-averaged internal energy,
\begin{equation}
\frac{E}{J} = -\frac{1}{2t_{1/2}}\int_{(k-1)2t_{1/2}}^{k2t_{1/2}} dt 
\frac{1}{L^2}\sum_{\langle ij \rangle}s_{i}(t)s_{j}(t)\;.
 \label{eq:E}
\end{equation}

As previous studies indicate \cite{SIDE98,SIDE99,KORN00}, 
the DPT transition essentially depends on the 
competition between two time scales: the average lifetime of the metastable 
phase, $\langle \tau (T,H_{0}) \rangle$, 
and the half-period of the applied field, 
$t_{1/2}$. The metastable lifetime $\langle \tau \rangle$ 
is defined as the average time it 
takes the system to leave one of its two degenerate zero-field equilibrium 
states, when a field of magnitude $H_{0}$ opposite to the initial 
magnetization is applied. In practice the metastable lifetime is measured as 
the first-passage time to zero magnetization. 

It is well known that metastable Ising models decay by different mechanisms 
depending on the magnitude of the applied field $H_{0}$, the system size $L$, 
and the temperature $T$. Detailed discussions of these different decays 
regimes are found in Ref.~\cite{RIKV94A}. 
More recently it has also been shown that, contrary 
to some common beliefs, there is also a strong dependence on the specific 
stochastic dynamics \cite{PARK04,BUEN04B}. 
For the purpose of this study the temperature, the system sizes and $H_0$ are 
chosen such that the metastable phase decays by random homogeneous nucleation 
of many critical droplets of the stable phase, which grow and coalesce, the 
so-called multidroplet (MD) regime. The metastable lifetime in the MD regime 
is independent of the system size \cite{RIKV94A}.

\section{Monte Carlo simulations} 
\label{sec:MC} 
The numerical simulations reported in this work are performed on square 
lattices with $L$ between 64 and 256 at $T=0.8T_c$. The system is subjected to 
an square-wave field $H(t)$ of amplitude $H_{0}=0.3J$.
The metastable lifetime was measured to be $\tau = 145\pm1$~MCSS, 
almost twice as long as for the 
kinetic Ising model evolving according to the standard (hard) Glauber dynamics
under the same conditions \cite{SIDE98,SIDE99,ROBB07}. 
This is consistent with earlier observations of slow nucleation
\cite{PARK04} and interface growth \cite{RIKV03} with this dynamic. 

The system was initialized with all the spins up, and the square-wave external 
field started in the half-period in which $H=-H_0$. After the system relaxed, 
the magnetization and energy reached a limit cycle (except for thermal 
fluctuations), and all the period-averaged quantities became stationary 
stochastic processes. We discarded the first 2000 periods of the time series to 
exclude transients from the stationary-state averages.

\begin{figure}[ht] 
\includegraphics[angle=0,width=.50\textwidth]{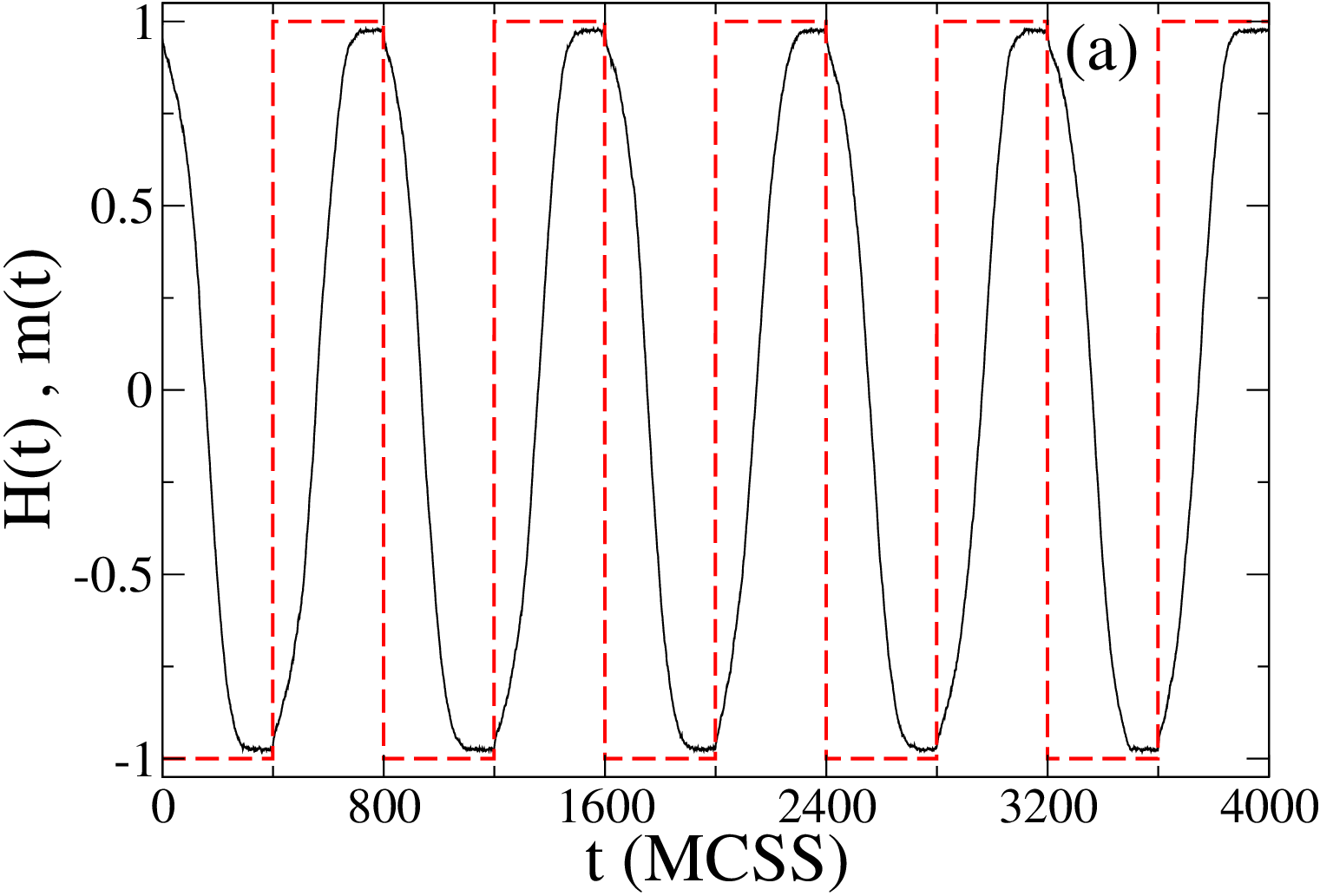}\\
\vspace{1truecm}
\includegraphics[angle=0,width=.50\textwidth]{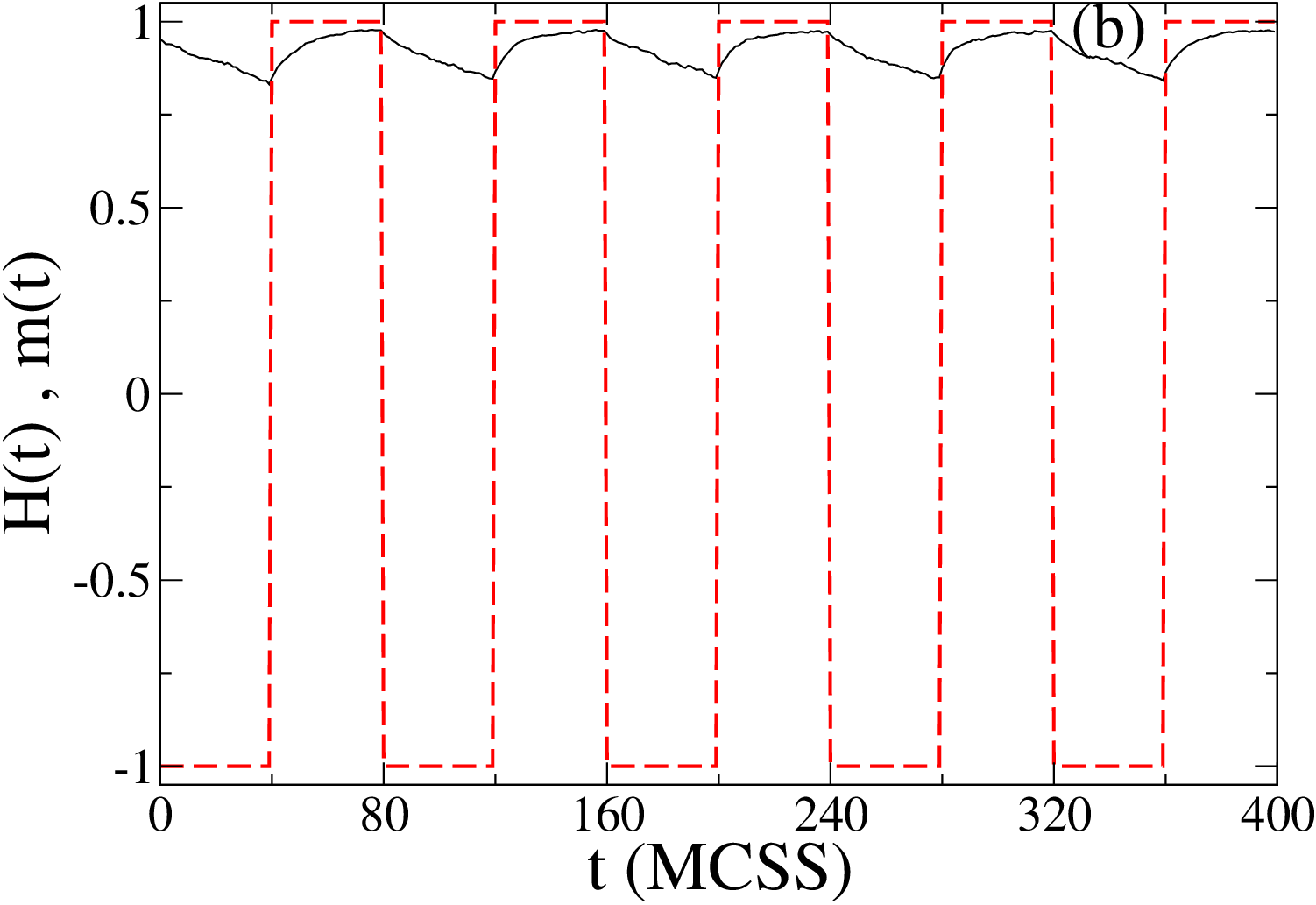}\\
\caption[]{
(Color online) Time series of the magnetization (solid curves) in the presence 
of a square-wave external field (dashed lines), for two values of the 
half-period $t_{1/2}$.
(a) $t_{1/2}=400$~MCSS, corresponding to a dynamically disordered phase.  
(b) $t_{1/2}=40$~MCSS, corresponding to a dynamically ordered phase.
The data were obtained for a system of size $L=128$ at $T=0.8T_c$ and field 
amplitude $H_{0}=0.3J$.
}
\label{fig:mag}
\end{figure}

\begin{figure}[ht]
\includegraphics[angle=0,width=.50\textwidth]{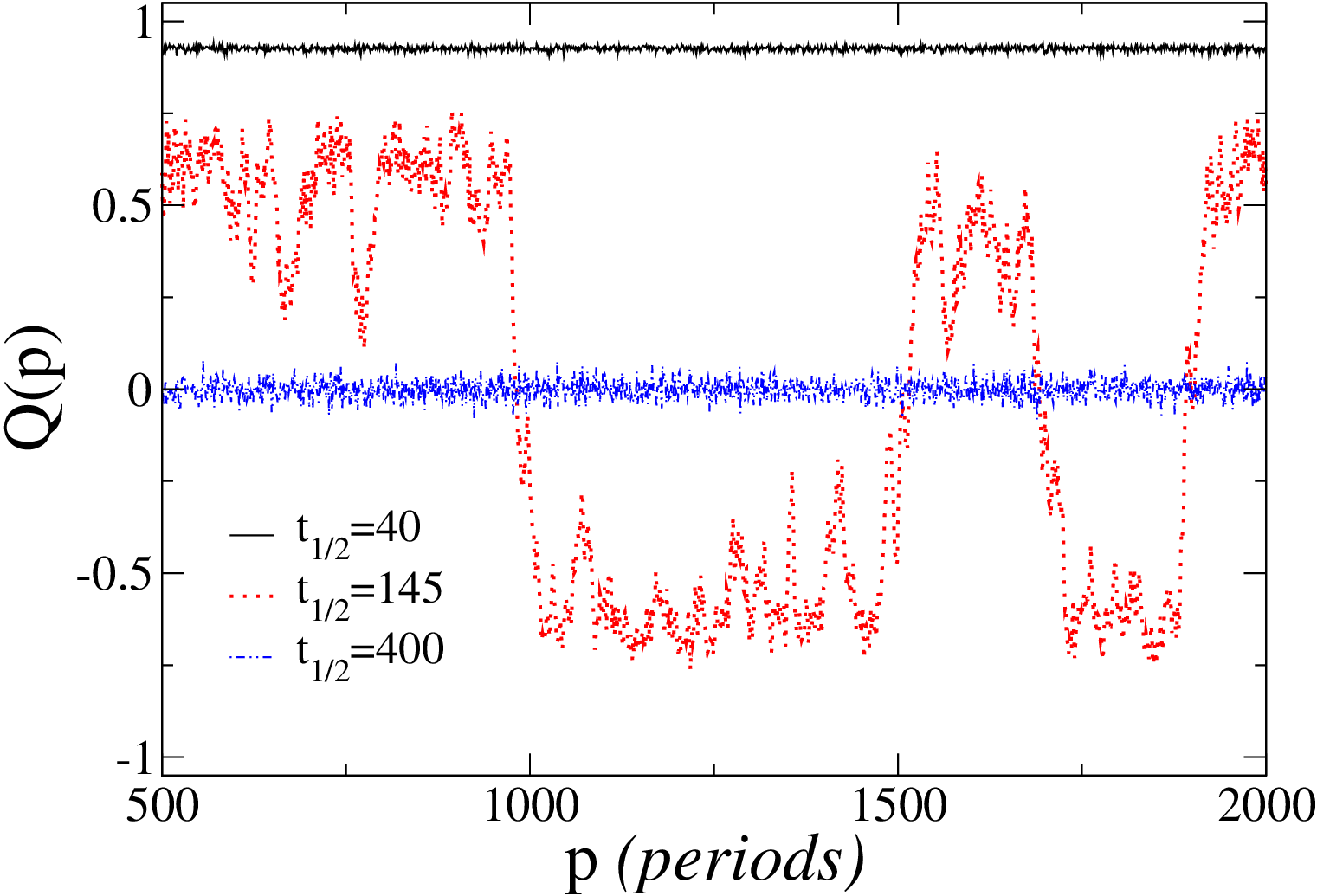}\\
\caption[]{
(Color online) Time series of the order parameter $Q$ for a system of size 
$L=128$ at $T=0.8T_{c}$ and $H_{0}=0.3J$. The horizontal trace near $Q=+1$ 
corresponds to a half-period of the field, $t_{1/2}=40$~MCSS,  
well into the dynamically ordered phase (see Fig.~\protect\ref{fig:mag}(b)). 
The strongly fluctuating trace corresponds to $t_{1/2}=145$~MCSS, 
very close to the DPT. The horizontal trace near $Q=0$ corresponds to 
$t_{1/2}=400$~MCSS, 
well into the dynamically disordered phase. (See
Fig.~\protect\ref{fig:mag}(a)).
}
\label{fig:q}
\end{figure}

The time evolution of the magnetization is shown in Fig.~\ref{fig:mag}.  
For slowly varying 
fields (Fig.~\ref{fig:mag}(a)), the magnetization follows the field, 
switching every 
half-period. In this region, $Q\approx 0$. For rapidly varying fields 
(Fig.~\ref{fig:mag}(b)), 
the magnetization does not have time to switch during a single half-period and 
remains nearly constant for many successive field cycles. As a result, 
the probability distribution of 
$Q$ becomes bimodal with two sharp peaks near the system's spontaneous 
equilibrium magnetization, $\pm m_{\rm sp}(T)$, 
corresponding to the broken symmetry of the hysteresis loop. The 
transition between these two regimes is characterized by large fluctuations in 
$Q$. This behavior of the time series $Q_k$, shown in Fig.~\ref{fig:q}, 
is a clear indication of the 
existence of a dynamical phase transition between a disordered dynamic 
phase (the region where $Q\approx 0$), and an ordered dynamic phase 
(where $Q\ne0$). Notice that the transition occurs at a critical value 
$\Theta_{\rm c} = t_{1/2}^{\rm c} / \tau$ that is very close to unity, 
the value at which the half-period of the external field is equal to the 
metastable lifetime of the system. 
To further explore the nature of the DPT, we perform a finite-size 
scaling analysis of the simulation data.

\subsection{Finite-size scaling}
\label{sec:FSS}

Previous studies indicate that although scaling laws and finite-size scaling 
are tools designed for equilibrium systems with a well known Hamiltonian, they 
can be successfully applied to far-from-equilibrium systems like the one we are 
analyzing here \cite{SIDE98,SIDE99,KORN00,MACH05,ROBB07}. 

Since for finite systems in the dynamically ordered phase the probability 
distribution of the order parameter is bimodal, in order to capture symmetry 
breaking, the order parameter is better defined as the average norm of $Q$, 
i.e., $\langle |Q| \rangle$. 
To characterize and quantify this transition by using finite-size scaling we 
must define quantities analogous to the susceptibility with respect to
the field conjugate to the order parameter in equilibrium systems. 
The scaled variance of the dynamic order parameter, 
\begin{equation} 
\chi_{L}^Q=L^{2}(\langle Q^{2}\rangle_{L}-\langle |Q| \rangle_{L}^2) \;.
\label{eq:chiQ}
\end{equation}
has long been used as a proxy for the non-equilibrium susceptibility. 
A fluctuation-dissipation relation was recently demonstrated, 
which justifies this practice by connecting $\chi_{L}^Q$ to the
susceptibility with respect to an applied bias field for a
two-dimensional kinetic Ising model
evolving under the standard Glauber dynamic \cite{ROBB07}. 

\begin{figure}[ht] 
\includegraphics[angle=0,width=.6\textwidth]{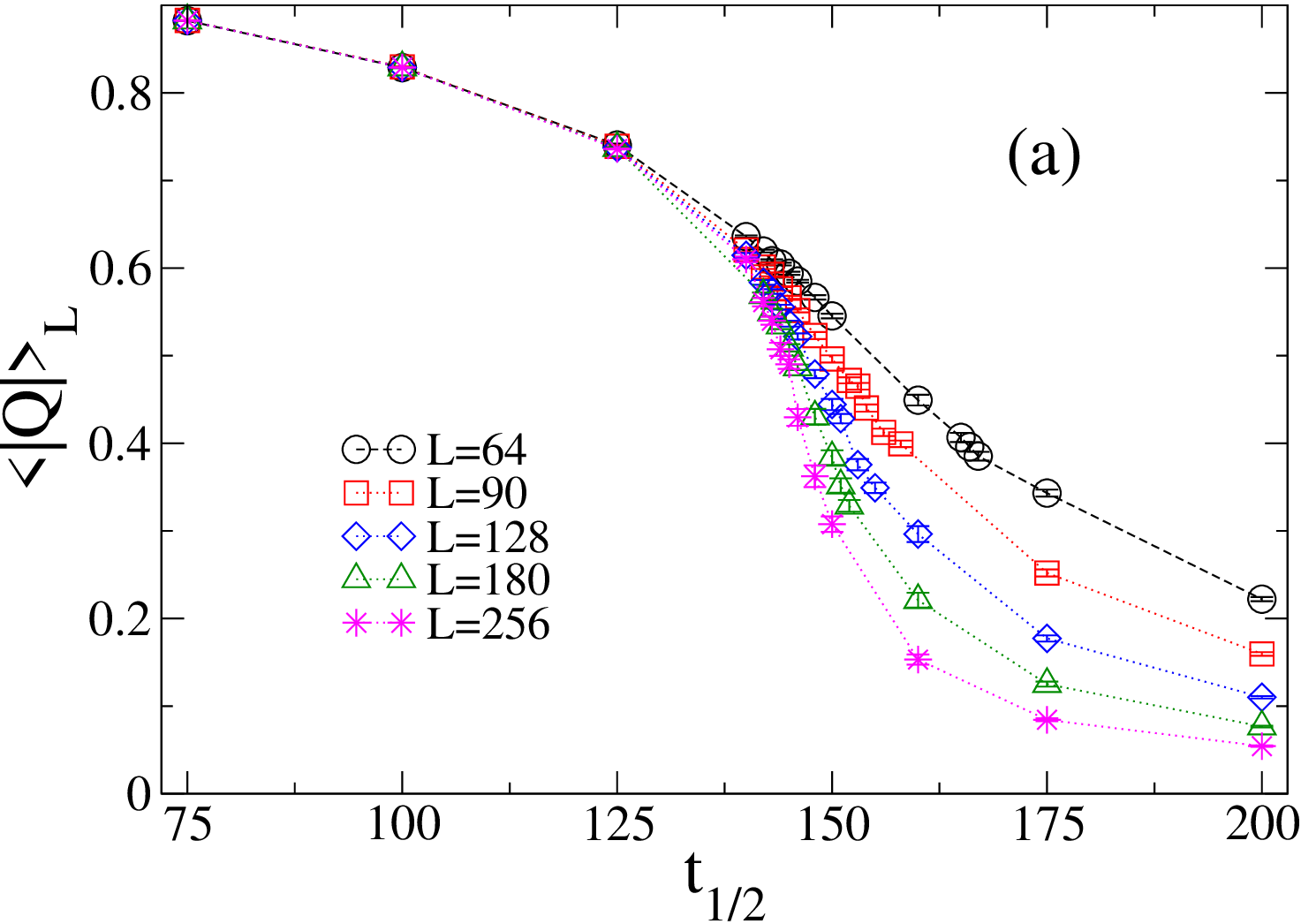}\\
\vspace{1.5truecm}
\includegraphics[angle=0,width=.6\textwidth]{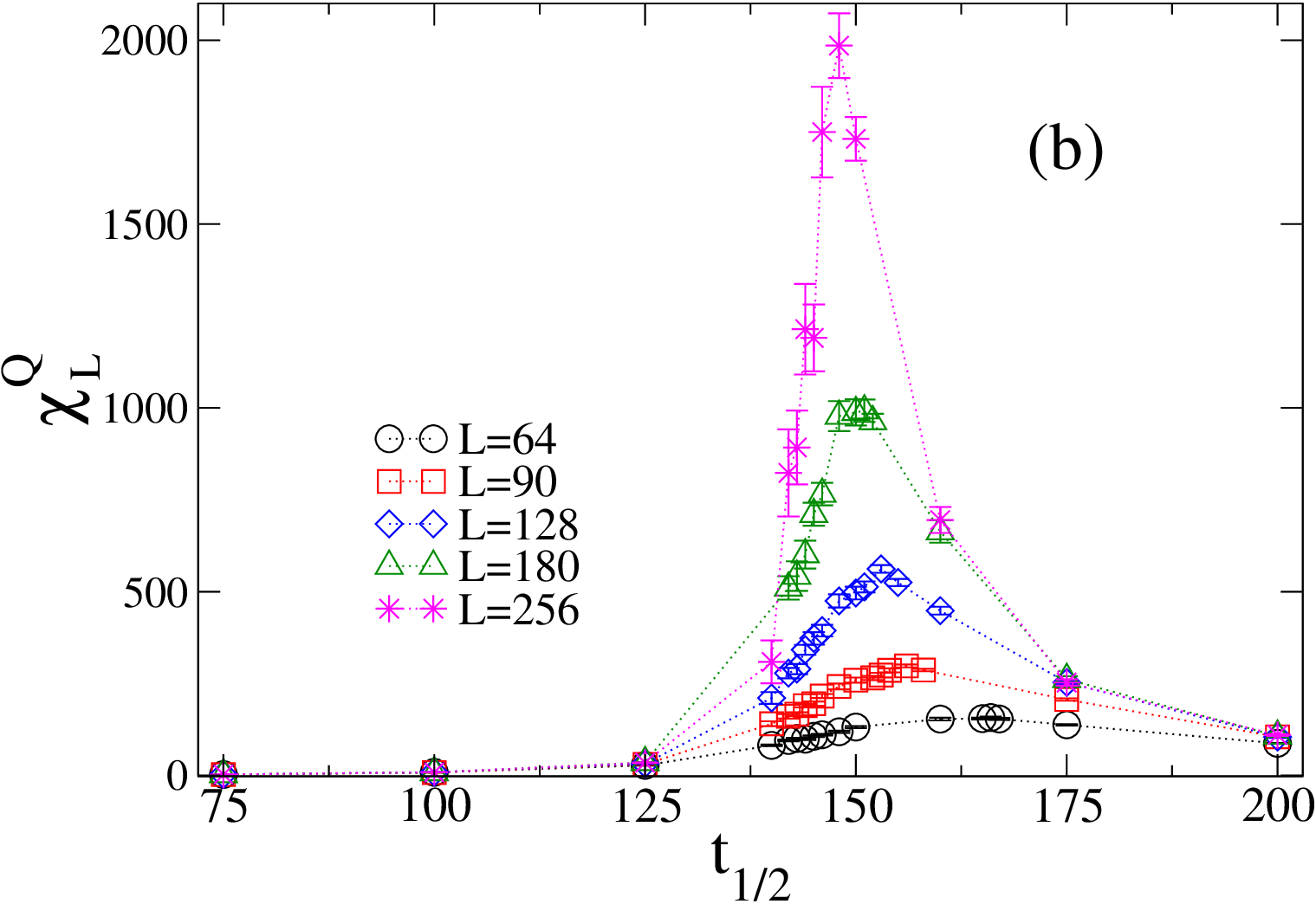}
\caption[]{(Color online)
Dependence on the half-period $t_{1/2}$  
of the order parameter $\langle |Q| \rangle$ (a), and of its scaled 
variance $\chi_{L}^Q$ (b), shown for 
various system sizes, $L$. All the results correspond to $T=0.8T_{c}$ and 
$H_{0}=0.3J$.
}
\label{fig:qn}
\end{figure}

In Fig.~\ref{fig:qn} we present the finite-size behavior of the order 
parameter and its fluctuations. Fig.~\ref{fig:qn}(a) 
shows that this dynamic
order parameter goes from unity to zero as $t_{1/2}$ increases, 
showing a sharp change around $t_{1/2}^c$,  characterized by the peak in 
$\chi_{L}^Q$ shown in Fig.~\ref{fig:qn}(b). 
The absence of finite-size effects below the 
critical point is the signature of the existence of a divergent length scale. 
The height and the location of the maximum in $\chi_{L}^Q$ change with $L$.

\begin{figure}[ht]
\includegraphics[angle=0,width=.45\textwidth]{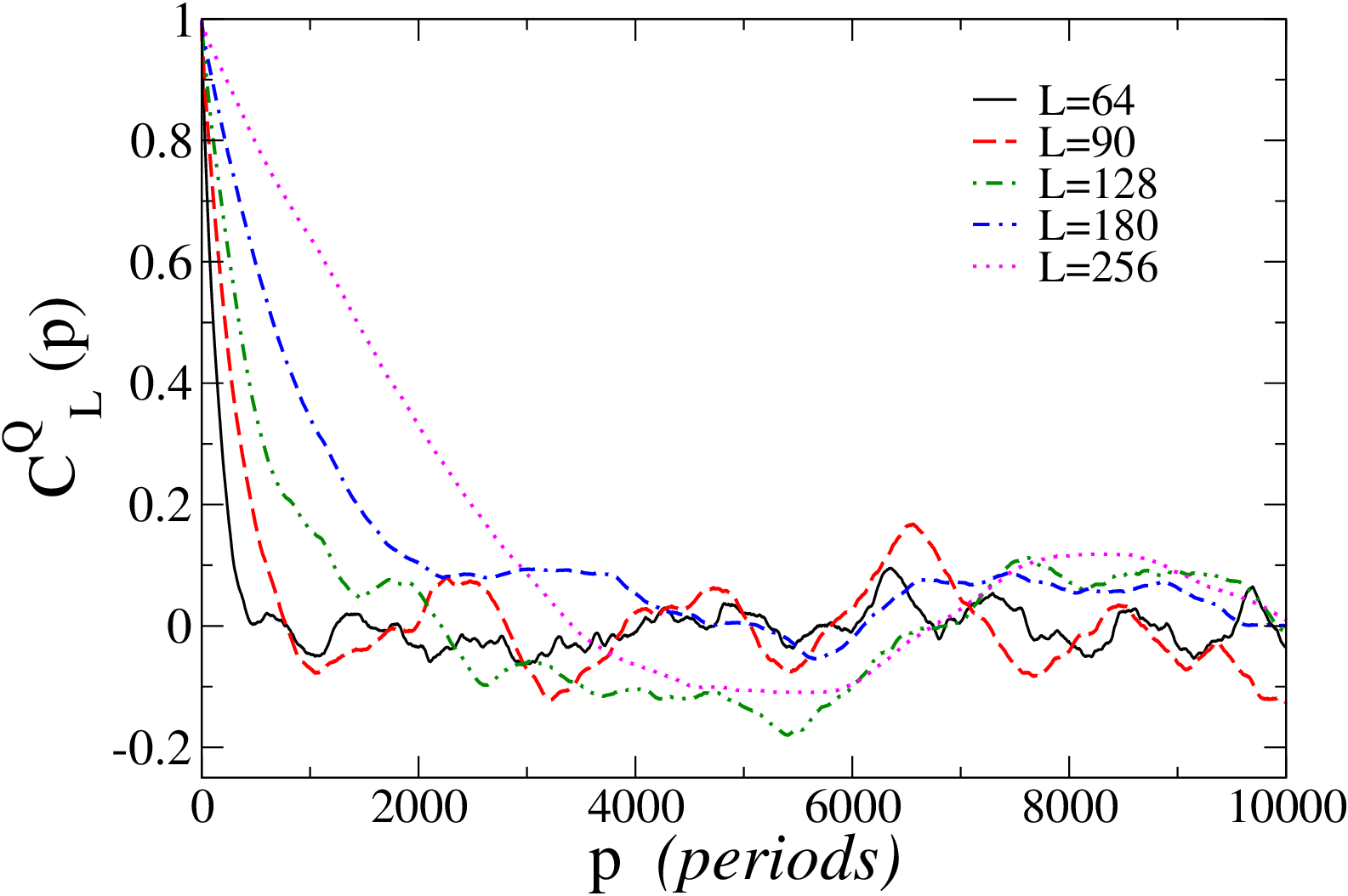}\\
\caption[]{
(Color online) Normalized autocorrelation function for the order parameter 
$Q$ for $t_{1/2}=145$~MCSS at $T=0.8T_c$ and $H_{0}=0.3J$, shown for
different values of $L$.
}
\label{fig:corr}
\end{figure}

In Fig.~\ref{fig:corr} we show the normalized time-autocorrelation 
function of the order parameter, defined as
\begin{equation}
 C_{L}^{Q}(p)=\frac{\langle Q(i)Q(i+p) \rangle-
 \langle Q(i)\rangle^{2}}{\langle Q(i)^{2}\rangle-\langle Q(i)\rangle^{2}}
.
\end{equation}
The increasing correlation times with increasing system sizes are 
evidence of the critical slowing down of the system, and provides 
further support for the existence of a DPT.

\begin{figure}[ht] 
\includegraphics[angle=0,width=.45\textwidth]{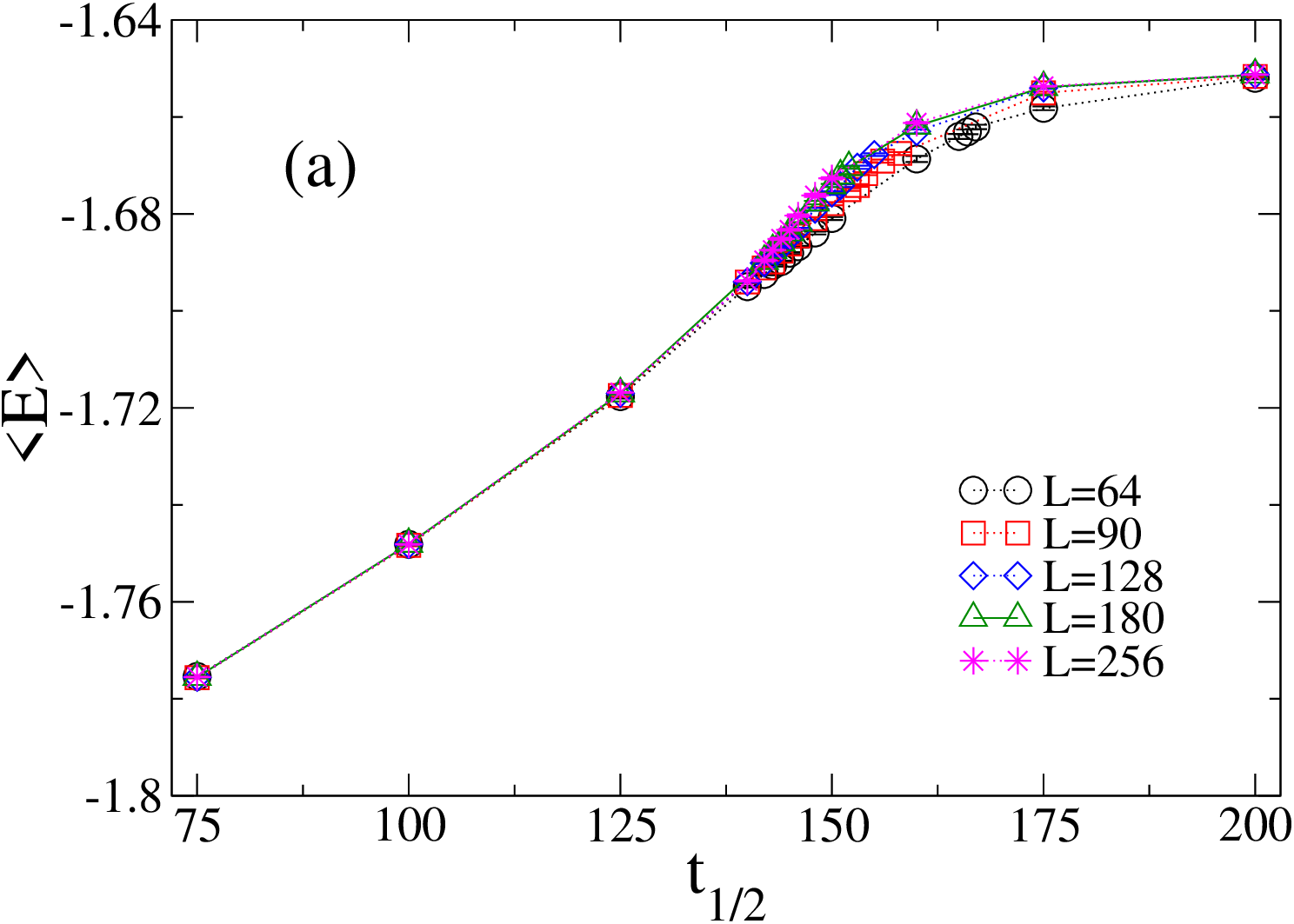}\\
\vspace{1truecm}
\includegraphics[angle=0,width=.45\textwidth]{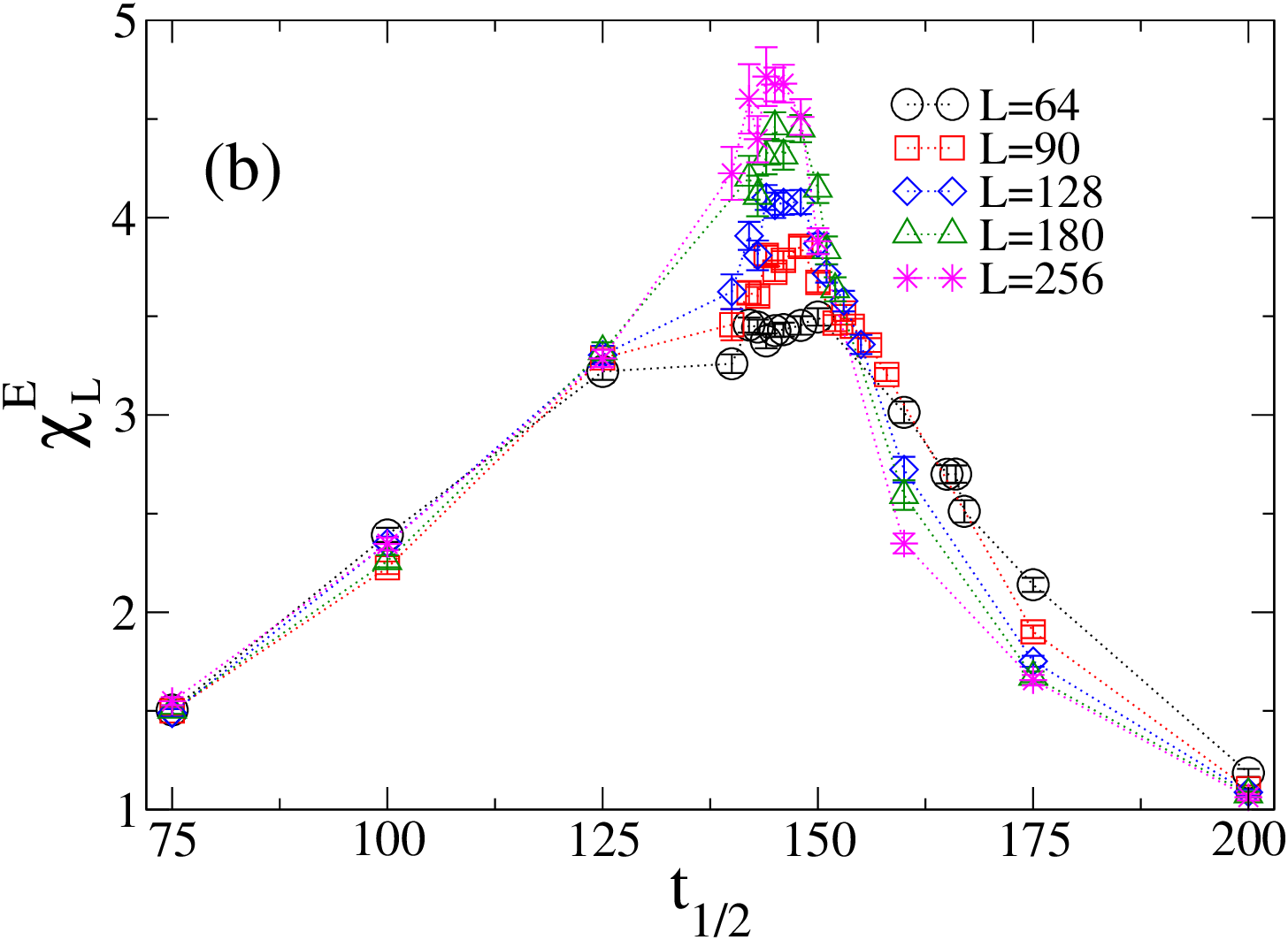}
\caption[]{(Color online)
Dependence on the half-period $t_{1/2}$ 
of the period-averaged internal energy $\langle E \rangle$ (a), and 
its scaled variance $\chi_{L}^E$ (b) 
for various system sizes, $L$. All the results correspond to $T=0.8T_{c}$ 
and $H_{0}=0.3J$.
}
\label{fig:energ}
\end{figure}

We also measured the period-averaged internal energy, Eq.~(\ref{eq:E}), 
and its scaled variance
\begin{equation} 
\chi_{L}^E=L^{2}(\langle E^{2} \rangle_{L}- \langle E\rangle_{L}^2) \;.
\label{eq:chiE}
\end{equation}
Both quantities are shown in Fig.~\ref{fig:energ}. Again, in the absence 
of a fluctuation-dissipation relation, we use the scaled variance as a
proxy for the analog of the equilibrium heat capacity.
The correlation time was used to estimate the proper sampling interval
for estimating the fluctuation measures and their error bars
as described in Ref.~\cite{LAND00}.

\begin{figure}[ht] 
\includegraphics[angle=0,width=.50\textwidth]{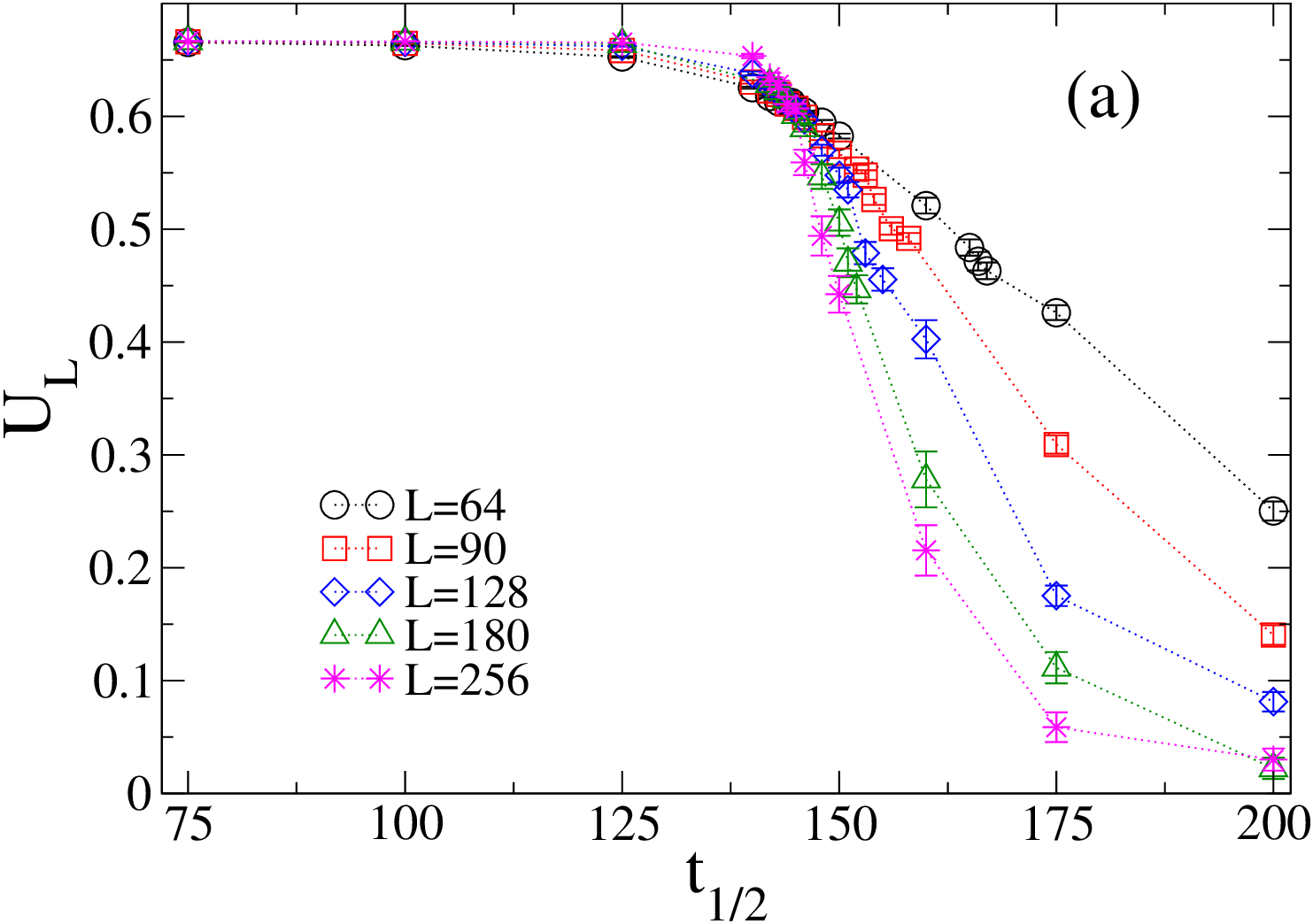}\\
\vspace{1truecm}
\includegraphics[angle=0,width=.50\textwidth]{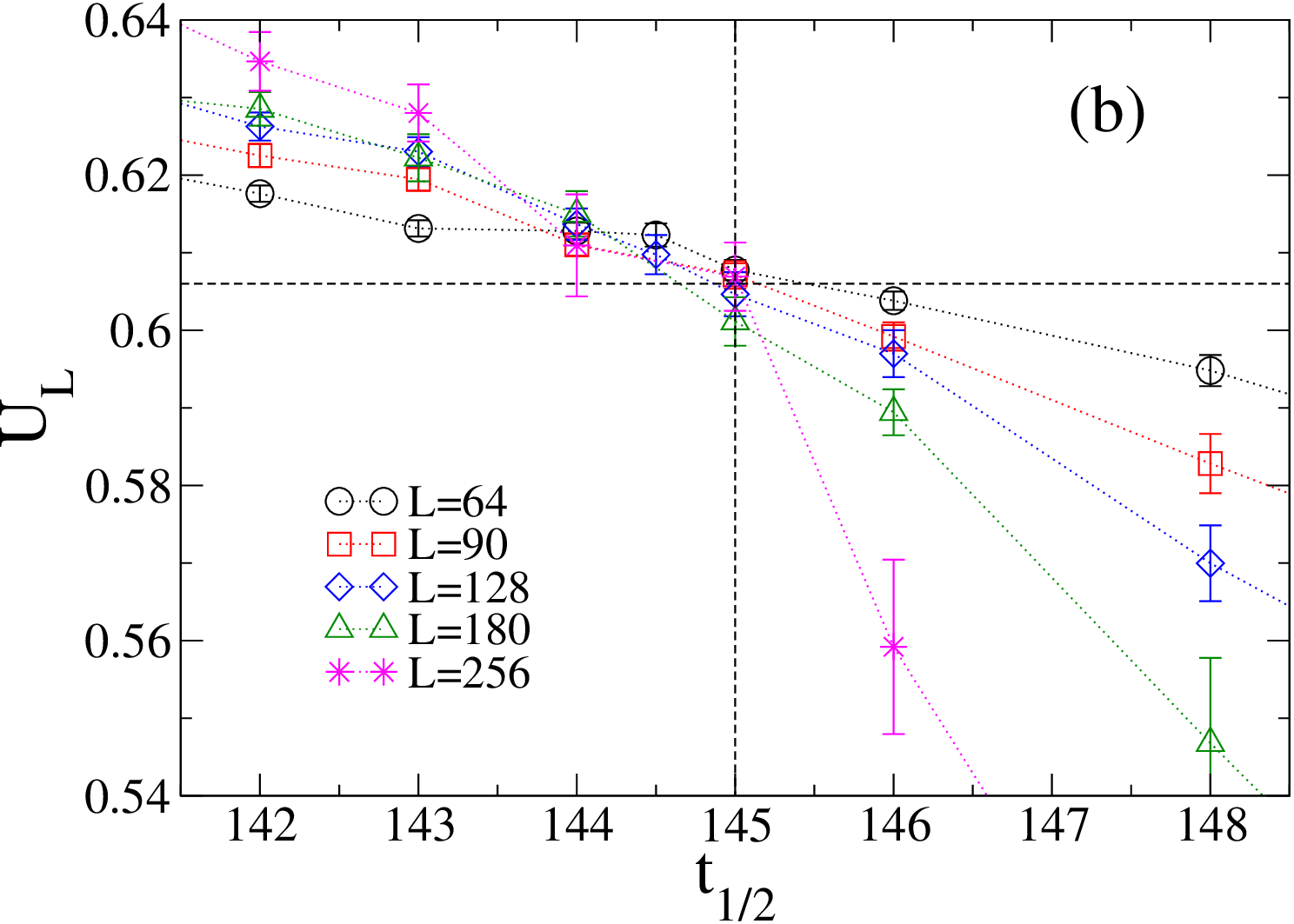}
\caption[]{(Color online)
(a) Dependence of the fourth-order cumulant $U_L$  on the half-period  
$t_{1/2}$, shown for various system sizes, $L$. 
(b) Enlargement of the region around 
the cumulant crossing. The horizontal and vertical dashed lines indicate 
the fixed point value $U^{*} \approx 0.606$ and the critical half-period, 
$t_{1/2}^{\rm c}=145$~MCSS, 
respectively.  All the results correspond to $T=0.8T_{c}$ and $H_{0}=0.3J$.
}
\label{fig:cum}
\end{figure}

It is very difficult to locate with precision the maxima of 
$\chi_{L}^Q$ and $\chi_{L}^E$ for the individual finite system 
sizes. A more accurate estimation of the critical point at which the 
transition occurs in an $\it{infinite}$ system can be obtained from the 
fourth-order cumulant intersection method. In Fig.~\ref{fig:cum} 
we plot the fourth-order cumulant $U_L$ defined as \cite{LAND00}
\begin{equation}
U_L=1-\frac{\langle Q^{4} \rangle_{L}}{3 \langle Q^{2}\rangle^{2}_{L}}
\label{eq:UL}
\end{equation}
as a function of $t_{1/2}$ for several system sizes. Our estimate is  
$t_{1/2}^{\rm c}=(145 \pm 1)$~MCSS, 
with a fixed-point value  $U^{*}=0.606\pm0.004$ for the cumulant 
(Fig.~\ref{fig:cum}(b)). 
The latter is consistent with the universal value for the
two-dimensional equilibrium Ising model, $U^* = 0.6106901(5)$ \cite{KAMI93}. 

\begin{figure}[ht] 
\includegraphics[angle=0,width=.50\textwidth]{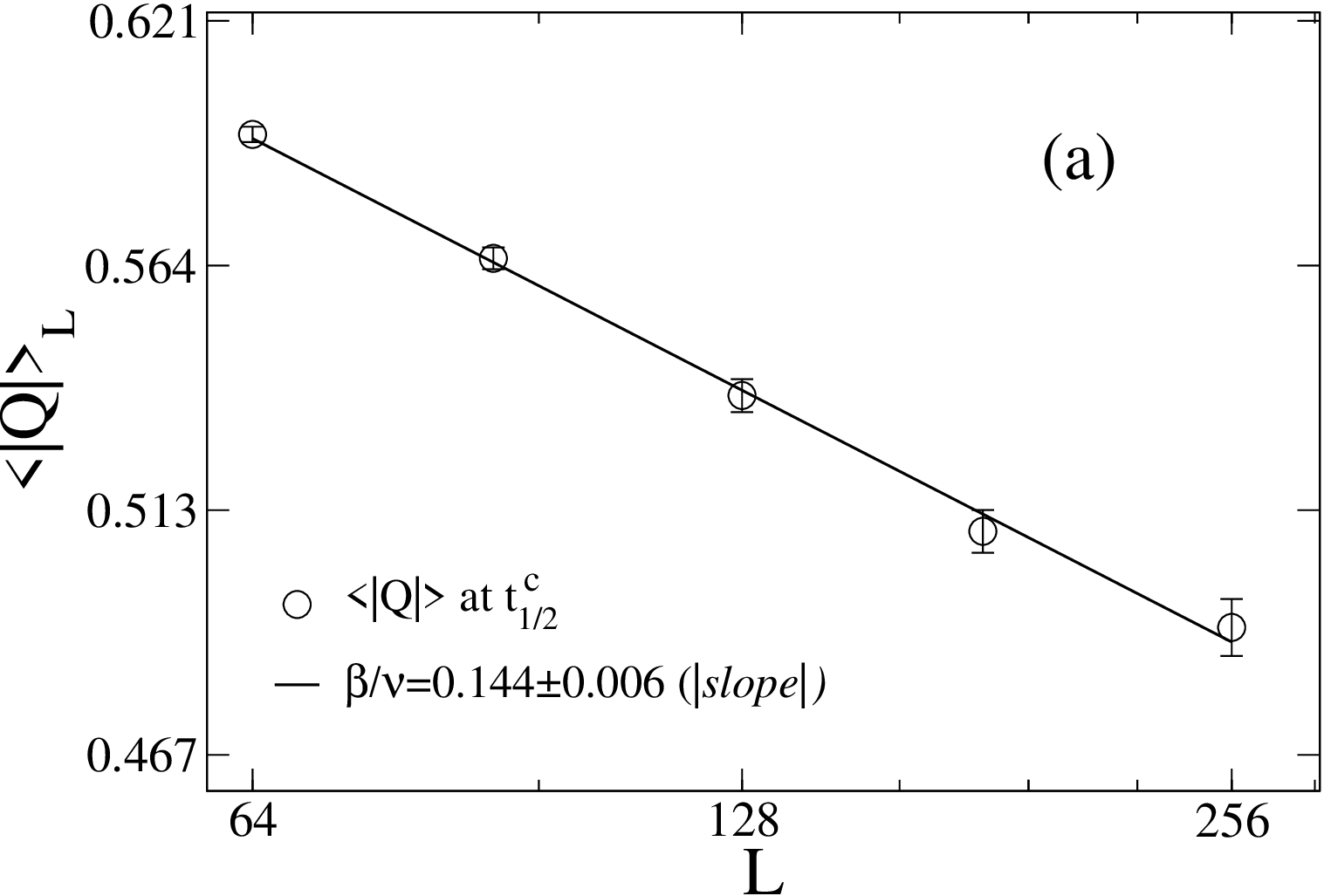}\\
\vspace{1truecm}
\includegraphics[angle=0,width=.50\textwidth]{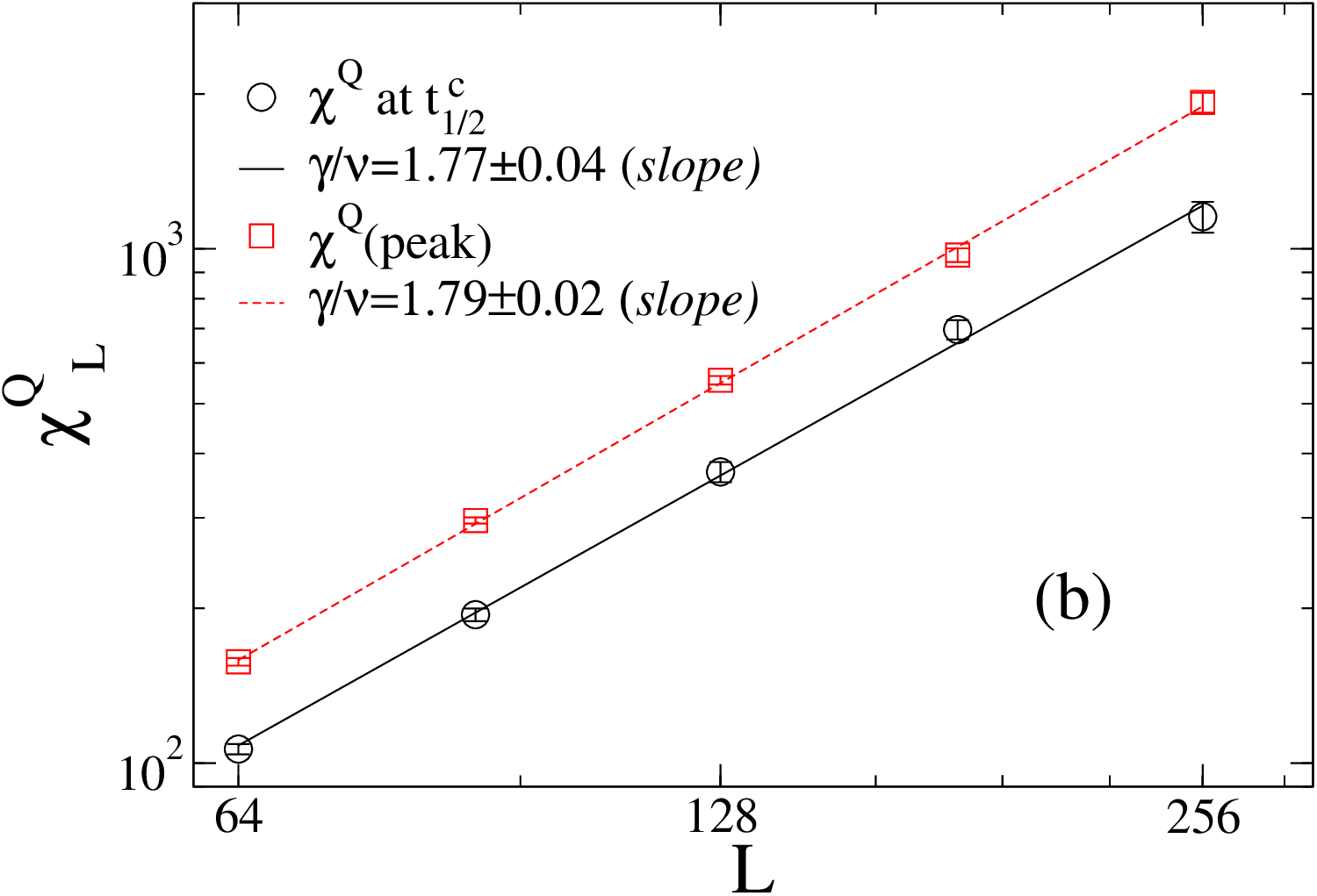}\\
\vspace{1truecm}
\includegraphics[angle=0,width=.50\textwidth]{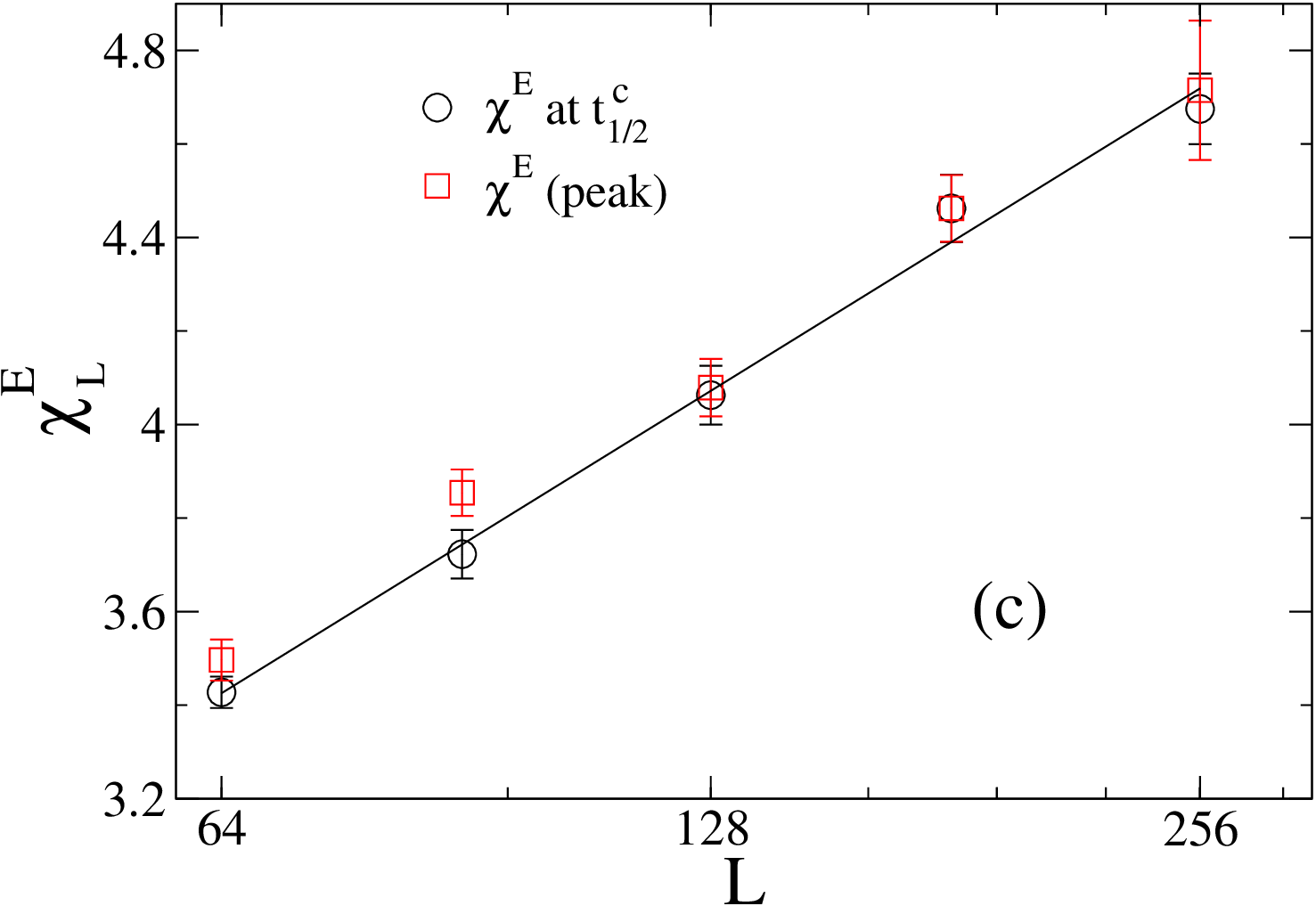}\\
\caption[]{(Color online)
Critical exponent estimates from the scaling relations. The symbols represent 
the MC data, the straight lines are weighted least-square fits. 
(a)Calculating $\beta/\nu$ from Eq.~(\ref{eq:QL}) 
calculating  $\gamma/\nu$ from Eq.~(\ref{eq:chiQL}). 
(c) The 
logarithmic divergence of the period-averaged energy fluctuations, based on 
Eq.~(\ref{eq:chiEL}). From data at $t_{1/2}^{\rm c}=145$~MCSS, 
circles, at the peaks, squares. 
}
\label{fig:expo}
\end{figure}

Finite size-scaling theory for equilibrium systems \cite{PRIV84,PRIV90} 
predicts the following scaling forms at the critical point,
\begin{equation}
\langle |Q| \rangle_{L} \propto L^{-\beta/\nu},
\label{eq:QL}
\end{equation}
\begin{equation}
\chi_{L}^Q \propto L^{\gamma/\nu},
\label{eq:chiQL}
\end{equation}
which are also applicable to the far-from-equilibrium DPT 
\cite{SIDE98,SIDE99,KORN00,ROBB07}. 
If the specific-heat 
critical exponent $\alpha=0$, as it is for the equilibrium Ising
universality class, then we also expect the logarithmic divergence, 
\begin{equation}
 \chi_{L}^{E} \propto A +B \ln (L)\;. 
\label{eq:chiEL}
\end{equation}
These relations enable us to estimate the critical exponent ratios $\beta/\nu$ 
and $\gamma/\nu$ and verify the logarithmic divergence in the period-averaged 
internal energy fluctuations. In Fig.~\ref{fig:expo} 
we present the results obtained by 
plotting the logarithm of $\langle |Q| \rangle_L$ (Fig.~\ref{fig:expo}(a)), 
the logarithm of $\chi_{L}^Q$ 
(Fig.~\ref{fig:expo}(b)), 
and $\chi_{L}^E$ (Fig.~\ref{fig:expo}(c)),  in term of the logarithm of $L$ at 
$t_{1/2}^{\rm c}$. 
We also plot the peak of the fluctuations, $\chi_{L}^Q({\rm Peak})$ 
(Fig.~\ref{fig:expo}(b)), and $\chi_{L}^E({\rm Peak})$ 
(Fig.~\ref{fig:expo}(c)), since they asymptotically should 
follow the same scaling laws. After fitting the data with a weighted, linear 
least-squares algorithm, our estimates for the critical exponents are: 
$\beta/\nu=1.44 \pm 0.06$, $\gamma/\nu=1.77 \pm 0.04$ 
(from the data at $t_{1/2}^{\rm c}$
), the data from $\chi_{L}^Q(Peak)$ gives $\gamma/\nu=1.79\pm.02$ which 
agree within statistical error. 
Also, the straight line in Fig.~\ref{fig:expo}(c) gives evidence of
the logarithmic divergence of $\chi_{L}^E$ at the critical point. 
These results, together with our estimate for $U^*$, give strong support to the
hypothesis that the DPT observed is in the same universality class of 
the {\it{equilibrium}} two-dimensional Ising model.

\section{Discussion and Conclusions}
\label{sec:DISC}
In this paper we have studied the dynamical response of a two-dimensional 
Ising model exposed to a square-wave oscillating external field. The system 
evolves under the so-called soft Glauber dynamic. In previous works it was 
established that, in the field and temperature regions when the metastable 
decay occurs via a multidroplet mechanism, this system evolving under a 
standard (hard) Glauber dynamics undergoes a continuous phase transition, with 
critical exponent ratios consistent with the equilibrium Ising values. 
The aim of the present study was  to explore the universality of this 
far-from-equilibrium DPT with respect to the dynamics chosen 
to evolve the system.

Our numerical results clearly indicate the existence of a DPT in the 
multidroplet regime. The transition depends on the competition between 
two time scales: the half-period of the applied field and the metastable 
lifetime of the system. We found that the metastable lifetime of the system 
evolving under the soft Glauber dynamics is roughly twice that of  
the same system evolving under the standard 
Glauber dynamic. However, in both cases the 
transition occurs at a critical point where both times 
are of the same order of magnitude. 
If the half-period of the applied field increases 
much beyond the metastable lifetime, the system is 
in a dynamically disordered phase characterized by a vanishing 
dynamic order parameter. 
A study of the autocorrelation function of the order parameter at the 
critical point provides evidence of critical slowing down, showing 
increasing correlation times with increasing system sizes.

We applied the machinery of finite-size scaling, originally developed
for equilibrium phase 
transitions, to estimate the critical point and the  
critical exponent ratios $\beta/\nu$ and $\gamma/\nu$ for system sizes 
between $64$ and $256$ at $T=0.8T_c$ and $H_{0}=0.3J$. Our estimates 
are $\beta/\nu \approx 1.44 \pm 0.06$ and $\gamma/\nu \approx 1.77 \pm 0.04$. 
These values are close to those of the two-dimensional equilibrium 
Ising model: $\beta/\nu=1/8=0.125$, $\gamma/\nu=7/4=1.75$. Furthermore, our 
data strongly indicate a slow logarithmic divergence with $L$
of the period-averaged energy fluctuations, consistent with the
equilibrium Ising exponent $\alpha = 0$.
The fixed-point value of the fourth-order cumulant, $U^*$,
is also close to its expected universal Ising value, near 0.611. 

This study provides further evidence of the universality class of the 
dynamic phase transition in kinetic Ising systems driven by an
oscillating field, extending its domain to systems that evolve under 
different stochastic dynamics that lead to interfaces with significantly
different structures on the nanoscale.

\section*{Acknowledgments}
\label{sec:ACK}

G.~M.~B.\ gratefully acknowledges many useful discussions with V.~M.~Kenkre and the hospitality of the 
Consortium of the Americas for Interdisciplinary Science at the University of New
Mexico, and P.~A.~R.\ that of the Department of Physics of The
University of Tokyo. Work at Florida State University was supported in
part by NSF Grants No.\ DMR-0444051 and DMR-0802288, and work at the University of New Mexico
was supported in part by NSF Grant No. \ INT-0336343. 

\clearpage





\begin{thebibliography}{43}
\expandafter\ifx\csname natexlab\endcsname\relax\def\natexlab#1{#1}\fi
\expandafter\ifx\csname bibnamefont\endcsname\relax
  \def\bibnamefont#1{#1}\fi
\expandafter\ifx\csname bibfnamefont\endcsname\relax
  \def\bibfnamefont#1{#1}\fi
\expandafter\ifx\csname url\endcsname\relax
  \def\url#1{\texttt{#1}}\fi
\expandafter\ifx\csname urlprefix\endcsname\relax\def\urlprefix{URL }\fi
\providecommand*{\bibinfo}[2]{#2}
\providecommand*{\eprint}[1]{#1}
\providecommand*{\url}[1]{#1}
\begingroup\makeatletter
 \@temptokena{%
  \expandafter\ifx\csname citenamefont\endcsname\relax
   \DeclareRobustCommand\citenamefont{\@firstofone}%
   \global\let\citenamefont\citenamefont
   \global\expandafter\let\csname citenamefont \expandafter\endcsname\csname
  citenamefont \endcsname
  \fi
 }\if@filesw\immediate\write\@auxout{\the\@temptokena}\fi
\expandafter\endgroup\the\@temptokena

\bibitem[{\citenamefont{Rikvold} \emph{et~al.}(1994)\citenamefont{Rikvold,
  Tomita, Miyashita, and Sides}}]{RIKV94A}
\bibinfo{author}{\bibfnamefont{P.~A.} \bibnamefont{Rikvold}},
  \bibinfo{author}{\bibfnamefont{H.}~\bibnamefont{Tomita}},
  \bibinfo{author}{\bibfnamefont{S.}~\bibnamefont{Miyashita}},
  \bibnamefont{and} \bibinfo{author}{\bibfnamefont{S.~W.} \bibnamefont{Sides}},
  \bibinfo{journal}{Phys.\ Rev.\ E} \textbf{\bibinfo{volume}{49}},
  \bibinfo{pages}{5080} (\bibinfo{year}{1994}).

\bibitem[{\citenamefont{Ramos} \emph{et~al.}(1999)\citenamefont{Ramos, Rikvold,
  and Novotny}}]{RAMO99}
\bibinfo{author}{\bibfnamefont{R.~A.} \bibnamefont{Ramos}},
  \bibinfo{author}{\bibfnamefont{P.~A.} \bibnamefont{Rikvold}},
  \bibnamefont{and} \bibinfo{author}{\bibfnamefont{M.~A.}
  \bibnamefont{Novotny}}, \bibinfo{journal}{Phys.\ Rev.\ B}
  \textbf{\bibinfo{volume}{59}}, \bibinfo{pages}{9053} (\bibinfo{year}{1999}).

\bibitem[{\citenamefont{Berthier}
  \emph{et~al.}(2004{\natexlab{a}})\citenamefont{Berthier, Legrand, Creuze,
  and Tetot}}]{BERT04}
\bibinfo{author}{\bibfnamefont{F.}~\bibnamefont{Berthier}},
  \bibinfo{author}{\bibfnamefont{B.}~\bibnamefont{Legrand}},
  \bibinfo{author}{\bibfnamefont{J.}~\bibnamefont{Creuze}}, \bibnamefont{and}
  \bibinfo{author}{\bibfnamefont{R.}~\bibnamefont{Tetot}},
  \bibinfo{journal}{J.\ Electroanal.\ Chem.} \textbf{\bibinfo{volume}{561}},
  \bibinfo{pages}{37} (\bibinfo{year}{2004}{\natexlab{a}}).

\bibitem[{\citenamefont{Berthier}
  \emph{et~al.}(2004{\natexlab{b}})\citenamefont{Berthier, Legrand, Creuze,
  and Tetot}}]{BERT04B}
\bibinfo{author}{\bibfnamefont{F.}~\bibnamefont{Berthier}},
  \bibinfo{author}{\bibfnamefont{B.}~\bibnamefont{Legrand}},
  \bibinfo{author}{\bibfnamefont{J.}~\bibnamefont{Creuze}}, \bibnamefont{and}
  \bibinfo{author}{\bibfnamefont{R.}~\bibnamefont{Tetot}},
  \bibinfo{journal}{J.\ Electroanal.\ Chem.} \textbf{\bibinfo{volume}{562}},
  \bibinfo{pages}{127} (\bibinfo{year}{2004}{\natexlab{b}}).

\bibitem{FRAN05}
S.~Frank, D.~E.\ Roberts, and P.~A.\ Rikvold, J.\ Chem.\ Phys. {\bf
122}, 064705 (2005). 

\bibitem[{\citenamefont{Frank and Rikvold}(2006)}]{FRAN06}
\bibinfo{author}{\bibfnamefont{S.}~\bibnamefont{Frank}} \bibnamefont{and}
  \bibinfo{author}{\bibfnamefont{P.~A.} \bibnamefont{Rikvold}},
  \bibinfo{journal}{Surf.\ Sci.} \textbf{\bibinfo{volume}{600}},
  \bibinfo{pages}{2470} (\bibinfo{year}{2006}).

\bibitem[{\citenamefont{Sides} \emph{et~al.}(1998)\citenamefont{Sides, Rikvold,
  and Novotny}}]{SIDE98}
\bibinfo{author}{\bibfnamefont{S.~W.} \bibnamefont{Sides}},
  \bibinfo{author}{\bibfnamefont{P.~A.} \bibnamefont{Rikvold}},
  \bibnamefont{and} \bibinfo{author}{\bibfnamefont{M.~A.}
  \bibnamefont{Novotny}}, \bibinfo{journal}{Phys.\ Rev.\ Lett.}
  \textbf{\bibinfo{volume}{81}}, \bibinfo{pages}{834} (\bibinfo{year}{1998}).

\bibitem[{\citenamefont{Sides} \emph{et~al.}(1999)\citenamefont{Sides, Rikvold,
  and Novotny}}]{SIDE99}
\bibinfo{author}{\bibfnamefont{S.~W.} \bibnamefont{Sides}},
  \bibinfo{author}{\bibfnamefont{P.~A.} \bibnamefont{Rikvold}},
  \bibnamefont{and} \bibinfo{author}{\bibfnamefont{M.~A.}
  \bibnamefont{Novotny}}, \bibinfo{journal}{Phys.\ Rev.\ E}
  \textbf{\bibinfo{volume}{59}}, \bibinfo{pages}{2710} (\bibinfo{year}{1999}).

\bibitem[{\citenamefont{Korniss} \emph{et~al.}(2001)\citenamefont{Korniss,
  White, Rikvold, and Novotny}}]{KORN00}
\bibinfo{author}{\bibfnamefont{G.}~\bibnamefont{Korniss}},
  \bibinfo{author}{\bibfnamefont{C.~J.} \bibnamefont{White}},
  \bibinfo{author}{\bibfnamefont{P.~A.} \bibnamefont{Rikvold}},
  \bibnamefont{and} \bibinfo{author}{\bibfnamefont{M.~A.}
  \bibnamefont{Novotny}}, \bibinfo{journal}{Phys.\ Rev.\ E}
  \textbf{\bibinfo{volume}{63}}, \bibinfo{pages}{016120}
  (\bibinfo{year}{2001}).

\bibitem[{\citenamefont{Richards} \emph{et~al.}(1995)\citenamefont{Richards,
  Sides, Novotny, and Rikvold}}]{RICH94}
\bibinfo{author}{\bibfnamefont{H.~L.} \bibnamefont{Richards}},
  \bibinfo{author}{\bibfnamefont{S.~W.} \bibnamefont{Sides}},
  \bibinfo{author}{\bibfnamefont{M.~A.} \bibnamefont{Novotny}},
  \bibnamefont{and} \bibinfo{author}{\bibfnamefont{P.~A.}
  \bibnamefont{Rikvold}}, \bibinfo{journal}{J.\ Magn.\ Magn.\ Mater.}
  \textbf{\bibinfo{volume}{150}}, \bibinfo{pages}{37} (\bibinfo{year}{1995}).

\bibitem[{\citenamefont{Novotny} \emph{et~al.}(2002)\citenamefont{Novotny,
  Brown, and Rikvold}}]{NOVO02A}
\bibinfo{author}{\bibfnamefont{M.~A.} \bibnamefont{Novotny}},
  \bibinfo{author}{\bibfnamefont{G.}~\bibnamefont{Brown}}, \bibnamefont{and}
  \bibinfo{author}{\bibfnamefont{P.~A.} \bibnamefont{Rikvold}},
  \bibinfo{journal}{J.\ Appl.\ Phys.} \textbf{\bibinfo{volume}{91}},
  \bibinfo{pages}{6908} (\bibinfo{year}{2002}).

\bibitem[{\citenamefont{Tom{\'e} and de~Oliveira}(1990)}]{TOME90}
\bibinfo{author}{\bibfnamefont{T.}~\bibnamefont{Tom{\'e}}} \bibnamefont{and}
  \bibinfo{author}{\bibfnamefont{M.~J.} \bibnamefont{de~Oliveira}},
  \bibinfo{journal}{Phys.\ Rev.\ A} \textbf{\bibinfo{volume}{41}},
  \bibinfo{pages}{4251} (\bibinfo{year}{1990}).

\bibitem[{\citenamefont{Mendes and Lage}(1991)}]{MEND91}
\bibinfo{author}{\bibfnamefont{J.~F.~F.} \bibnamefont{Mendes}}
  \bibnamefont{and} \bibinfo{author}{\bibfnamefont{E.~J.~S.}
  \bibnamefont{Lage}}, \bibinfo{journal}{J.\ Stat.\ Phys.}
  \textbf{\bibinfo{volume}{64}}, \bibinfo{pages}{653} (\bibinfo{year}{1991}).

\bibitem[{\citenamefont{Zimmer}(1993)}]{ZIMM93A}
\bibinfo{author}{\bibfnamefont{M.}~\bibnamefont{Zimmer}},
  \bibinfo{journal}{Phys.\ Rev.\ E} \textbf{\bibinfo{volume}{47}},
  \bibinfo{pages}{3950} (\bibinfo{year}{1993}).
  
\bibitem{BUEN98}
G.~M.\ Buend{\'\i}a and E.~Machado, Phys.\ Rev.\ E {\bf 58}, 1260 (1998).

\bibitem[{\citenamefont{Acharyya and Chakrabarti}(1995)}]{ACHA95}
\bibinfo{author}{\bibfnamefont{M.}~\bibnamefont{Acharyya}} \bibnamefont{and}
  \bibinfo{author}{\bibfnamefont{B.}~\bibnamefont{Chakrabarti}},
  \bibinfo{journal}{Phys.\ Rev.\ B} \textbf{\bibinfo{volume}{52}},
  \bibinfo{pages}{6550} (\bibinfo{year}{1995}).

\bibitem[{\citenamefont{Chakrabarti and Acharyya}(1999)}]{CHAK99}
\bibinfo{author}{\bibfnamefont{B.}~\bibnamefont{Chakrabarti}} \bibnamefont{and}
  \bibinfo{author}{\bibfnamefont{M.}~\bibnamefont{Acharyya}},
  \bibinfo{journal}{Rev.\ Mod.\ Phys.} \textbf{\bibinfo{volume}{71}},
  \bibinfo{pages}{847} (\bibinfo{year}{1999}).

\bibitem[{\citenamefont{Lo and Pelcovits}(1990)}]{LO90}
\bibinfo{author}{\bibfnamefont{W.~S.} \bibnamefont{Lo}} \bibnamefont{and}
  \bibinfo{author}{\bibfnamefont{R.~A.} \bibnamefont{Pelcovits}},
  \bibinfo{journal}{Phys.\ Rev.\ A} \textbf{\bibinfo{volume}{42}},
  \bibinfo{pages}{7471} (\bibinfo{year}{1990}).

\bibitem[{\citenamefont{Korniss} \emph{et~al.}(2002)\citenamefont{Korniss,
  Rikvold, and Novotny}}]{KORN02B}
\bibinfo{author}{\bibfnamefont{G.}~\bibnamefont{Korniss}},
  \bibinfo{author}{\bibfnamefont{P.~A.} \bibnamefont{Rikvold}},
  \bibnamefont{and} \bibinfo{author}{\bibfnamefont{M.~A.}
  \bibnamefont{Novotny}}, \bibinfo{journal}{Phys.\ Rev.\ E}
  \textbf{\bibinfo{volume}{66}}, \bibinfo{pages}{056127}
  (\bibinfo{year}{2002}).

\bibitem[{\citenamefont{Robb} \emph{et~al.}(2007)\citenamefont{Robb, Rikvold,
  Berger, and Novotny}}]{ROBB07}
\bibinfo{author}{\bibfnamefont{D.~T.} \bibnamefont{Robb}},
  \bibinfo{author}{\bibfnamefont{P.~A.} \bibnamefont{Rikvold}},
  \bibinfo{author}{\bibfnamefont{A.}~\bibnamefont{Berger}}, \bibnamefont{and}
  \bibinfo{author}{\bibfnamefont{M.~A.} \bibnamefont{Novotny}},
  \bibinfo{journal}{Phys.\ Rev.\ E} \textbf{\bibinfo{volume}{76}},
  \bibinfo{pages}{021124} (\bibinfo{year}{2007}).

\bibitem[{\citenamefont{Fujisaka} \emph{et~al.}(2001)\citenamefont{Fujisaka,
  Tutu, and Rikvold}}]{FUJI01}
\bibinfo{author}{\bibfnamefont{H.}~\bibnamefont{Fujisaka}},
  \bibinfo{author}{\bibfnamefont{H.}~\bibnamefont{Tutu}}, \bibnamefont{and}
  \bibinfo{author}{\bibfnamefont{P.~A.} \bibnamefont{Rikvold}},
  \bibinfo{journal}{Phys.\ Rev.\ E} \textbf{\bibinfo{volume}{63}},
  \bibinfo{pages}{036109} (\bibinfo{year}{2001}); \bibinfo{note}{erratum: 
  {\bf 63}, 059903 (2001).}

\bibitem[{\citenamefont{Tutu and Fujiwara}(2004)}]{TUTU04}
\bibinfo{author}{\bibfnamefont{H.}~\bibnamefont{Tutu}} \bibnamefont{and}
  \bibinfo{author}{\bibfnamefont{N.}~\bibnamefont{Fujiwara}},
  \bibinfo{journal}{J.\ Phys.\ Soc.\ Jpn.} \textbf{\bibinfo{volume}{73}},
  \bibinfo{pages}{2680} (\bibinfo{year}{2004}).

\bibitem[{\citenamefont{Meilikhov}(2004)}]{MEIL04}
\bibinfo{author}{\bibfnamefont{E.~Z.} \bibnamefont{Meilikhov}},
  \bibinfo{journal}{JETP Lett.} \textbf{\bibinfo{volume}{79}},
  \bibinfo{pages}{620} (\bibinfo{year}{2004}).

\bibitem[{\citenamefont{Dutta}(2004)}]{DUTT04}
\bibinfo{author}{\bibfnamefont{S.~B.} \bibnamefont{Dutta}},
  \bibinfo{journal}{Phys.\ Rev.\ E} \textbf{\bibinfo{volume}{69}},
  \bibinfo{pages}{066115} (\bibinfo{year}{2004}).

\bibitem[{\citenamefont{Robb} \emph{et~al.}(2008)\citenamefont{Robb, Xu,
  Hellwig, Berger, Novotny, and Rikvold}}]{ROBB08}
\bibinfo{author}{\bibfnamefont{D.~T.} \bibnamefont{Robb}},
  \bibinfo{author}{\bibfnamefont{Y.~H.} \bibnamefont{Xu}},
  \bibinfo{author}{\bibfnamefont{A.}~\bibnamefont{Hellwig}},
  \bibinfo{author}{\bibfnamefont{A.}~\bibnamefont{Berger}},
  \bibinfo{author}{\bibfnamefont{M.~A.} \bibnamefont{Novotny}},
  \bibnamefont{and} \bibinfo{author}{\bibfnamefont{P.~A.}
  \bibnamefont{Rikvold}}  (\bibinfo{year}{2008}), \bibinfo{note}{e-print
  arXiv:0705.4454}.

\bibitem[{\citenamefont{Machado} \emph{et~al.}(2005)\citenamefont{Machado,
  Buend{\'\i}a, Rikvold, and Ziff}}]{MACH05}
\bibinfo{author}{\bibfnamefont{E.}~\bibnamefont{Machado}},
  \bibinfo{author}{\bibfnamefont{G.~M.} \bibnamefont{Buend{\'\i}a}},
  \bibinfo{author}{\bibfnamefont{P.~A.} \bibnamefont{Rikvold}},
  \bibnamefont{and} \bibinfo{author}{\bibfnamefont{R.~M.} \bibnamefont{Ziff}},
  \bibinfo{journal}{Phys.\ Rev.\ E} \textbf{\bibinfo{volume}{71}},
  \bibinfo{pages}{016120} (\bibinfo{year}{2005}).

\bibitem[{\citenamefont{Buend{\'\i}a}
  \emph{et~al.}(2006{\natexlab{a}})\citenamefont{Buend{\'\i}a, Machado, and
  Rikvold}}]{BUEN06C}
\bibinfo{author}{\bibfnamefont{G.~M.} \bibnamefont{Buend{\'\i}a}},
  \bibinfo{author}{\bibfnamefont{E.}~\bibnamefont{Machado}}, \bibnamefont{and}
  \bibinfo{author}{\bibfnamefont{P.~A.} \bibnamefont{Rikvold}},
  \bibinfo{journal}{J.\ Mol.\ Struct.: THEOCHEM}
  \textbf{\bibinfo{volume}{769}}, \bibinfo{pages}{189}
  (\bibinfo{year}{2006}{\natexlab{a}}).

\bibitem[{\citenamefont{Grinstein} \emph{et~al.}(1985)\citenamefont{Grinstein,
  Jayaprakash, and He}}]{GRIN85}
\bibinfo{author}{\bibfnamefont{G.}~\bibnamefont{Grinstein}},
  \bibinfo{author}{\bibfnamefont{C.}~\bibnamefont{Jayaprakash}},
  \bibnamefont{and} \bibinfo{author}{\bibfnamefont{Y.}~\bibnamefont{He}},
  \bibinfo{journal}{Phys.\ Rev.\ Lett.} \textbf{\bibinfo{volume}{55}},
  \bibinfo{pages}{2527} (\bibinfo{year}{1985}).

\bibitem[{\citenamefont{Bassler and Schmittmann}(1994)}]{BASS94}
\bibinfo{author}{\bibfnamefont{K.~E.} \bibnamefont{Bassler}} \bibnamefont{and}
  \bibinfo{author}{\bibfnamefont{B.}~\bibnamefont{Schmittmann}},
  \bibinfo{journal}{Phys.\ Rev.\ Lett.} \textbf{\bibinfo{volume}{73}},
  \bibinfo{pages}{3343} (\bibinfo{year}{1994}).

\bibitem[{\citenamefont{Glauber}(1963)}]{GLAU63}
\bibinfo{author}{\bibfnamefont{R.~J.} \bibnamefont{Glauber}},
  \bibinfo{journal}{J.\ Math.\ Phys.} \textbf{\bibinfo{volume}{4}},
  \bibinfo{pages}{294} (\bibinfo{year}{1963}).

\bibitem[{\citenamefont{Metropolis}
  \emph{et~al.}(1953)\citenamefont{Metropolis, Rosenbluth, Rosenbluth, Teller,
  and Teller}}]{METR53}
\bibinfo{author}{\bibfnamefont{N.}~\bibnamefont{Metropolis}},
  \bibinfo{author}{\bibfnamefont{A.~W.} \bibnamefont{Rosenbluth}},
  \bibinfo{author}{\bibfnamefont{M.~N.} \bibnamefont{Rosenbluth}},
  \bibinfo{author}{\bibfnamefont{A.~H.} \bibnamefont{Teller}},
  \bibnamefont{and} \bibinfo{author}{\bibfnamefont{E.}~\bibnamefont{Teller}},
  \bibinfo{journal}{J.\ Chem.\ Phys.} \textbf{\bibinfo{volume}{21}},
  \bibinfo{pages}{1087} (\bibinfo{year}{1953}).

\bibitem[{\citenamefont{Landau and Binder}(2000)}]{LAND00}
\bibinfo{author}{\bibfnamefont{D.~P.} \bibnamefont{Landau}} \bibnamefont{and}
  \bibinfo{author}{\bibfnamefont{K.}~\bibnamefont{Binder}},
  \emph{\bibinfo{title}{A Guide to Monte Carlo Simulations in Statistical
  Physics}} (\bibinfo{publisher}{Cambridge University Press},
  \bibinfo{address}{Cambridge}, \bibinfo{year}{2000}).

\bibitem[{\citenamefont{Gunton} \emph{et~al.}(1983)\citenamefont{Gunton,
  San~Miguel, and Sahni}}]{GUNT83B}
\bibinfo{author}{\bibfnamefont{J.~D.} \bibnamefont{Gunton}},
  \bibinfo{author}{\bibfnamefont{M.}~\bibnamefont{San~Miguel}},
  \bibnamefont{and} \bibinfo{author}{\bibfnamefont{P.~S.} \bibnamefont{Sahni}},
  in \emph{\bibinfo{booktitle}{Phase Transitions and Critical Phenomena,
  Vol.~8}}, edited by \bibinfo{editor}{\bibfnamefont{C.}~\bibnamefont{Domb}}
  \bibnamefont{and} \bibinfo{editor}{\bibfnamefont{J.~L.}
  \bibnamefont{Lebowitz}} (\bibinfo{publisher}{Academic},
  \bibinfo{address}{London}, \bibinfo{year}{1983}).

\bibitem[{\citenamefont{Park} \emph{et~al.}(2004)\citenamefont{Park, Rikvold,
  Buend{\'\i}a, and Novotny}}]{PARK04}
\bibinfo{author}{\bibfnamefont{K.}~\bibnamefont{Park}},
  \bibinfo{author}{\bibfnamefont{P.~A.} \bibnamefont{Rikvold}},
  \bibinfo{author}{\bibfnamefont{G.~M.} \bibnamefont{Buend{\'\i}a}},
  \bibnamefont{and} \bibinfo{author}{\bibfnamefont{M.~A.}
  \bibnamefont{Novotny}}, \bibinfo{journal}{Phys.\ Rev.\ Lett.}
  \textbf{\bibinfo{volume}{92}}, \bibinfo{pages}{015701}
  (\bibinfo{year}{2004}).

\bibitem[{\citenamefont{Buend\'{\i}a}
  \emph{et~al.}(2004)\citenamefont{Buend\'{\i}a, Rikvold, Park, and
  Novotny}}]{BUEN04B}
\bibinfo{author}{\bibfnamefont{G.~M.} \bibnamefont{Buend\'{\i}a}},
  \bibinfo{author}{\bibfnamefont{P.~A.} \bibnamefont{Rikvold}},
  \bibinfo{author}{\bibfnamefont{K.}~\bibnamefont{Park}}, \bibnamefont{and}
  \bibinfo{author}{\bibfnamefont{M.~A.} \bibnamefont{Novotny}},
  \bibinfo{journal}{J.\ Chem.\ Phys.} \textbf{\bibinfo{volume}{121}},
  \bibinfo{pages}{4193} (\bibinfo{year}{2004}).

\bibitem[{\citenamefont{Rikvold and Kolesik}(2002{\natexlab{a}})}]{RIKV02}
\bibinfo{author}{\bibfnamefont{P.~A.} \bibnamefont{Rikvold}} \bibnamefont{and}
  \bibinfo{author}{\bibfnamefont{M.}~\bibnamefont{Kolesik}},
  \bibinfo{journal}{J.\ Phys.\ A: Math.\ Gen.} \textbf{\bibinfo{volume}{35}},
  \bibinfo{pages}{L117} (\bibinfo{year}{2002}{\natexlab{a}}).

\bibitem[{\citenamefont{Rikvold and Kolesik}(2002{\natexlab{b}})}]{RIKV02B}
\bibinfo{author}{\bibfnamefont{P.~A.} \bibnamefont{Rikvold}} \bibnamefont{and}
  \bibinfo{author}{\bibfnamefont{M.}~\bibnamefont{Kolesik}},
  \bibinfo{journal}{Phys.\ Rev.\ E} \textbf{\bibinfo{volume}{66}},
  \bibinfo{pages}{066116} (\bibinfo{year}{2002}{\natexlab{b}}).

\bibitem[{\citenamefont{Rikvold and Kolesik}(2003)}]{RIKV03}
\bibinfo{author}{\bibfnamefont{P.~A.} \bibnamefont{Rikvold}} \bibnamefont{and}
  \bibinfo{author}{\bibfnamefont{M.}~\bibnamefont{Kolesik}},
  \bibinfo{journal}{Phys.\ Rev.\ E} \textbf{\bibinfo{volume}{67}},
  \bibinfo{pages}{066113} (\bibinfo{year}{2003}).

\bibitem[{\citenamefont{Buend{\'\i}a}
  \emph{et~al.}(2006{\natexlab{b}})\citenamefont{Buend{\'\i}a, Rikvold, and
  Kolesik}}]{BUEN06}
\bibinfo{author}{\bibfnamefont{G.~M.} \bibnamefont{Buend{\'\i}a}},
  \bibinfo{author}{\bibfnamefont{P.~A.} \bibnamefont{Rikvold}},
  \bibnamefont{and} \bibinfo{author}{\bibfnamefont{M.}~\bibnamefont{Kolesik}},
  \bibinfo{journal}{Phys.\ Rev.\ B} \textbf{\bibinfo{volume}{73}},
  \bibinfo{pages}{045437} (\bibinfo{year}{2006}{\natexlab{b}}).

\bibitem[{\citenamefont{Buend{\'\i}a}
  \emph{et~al.}(2006{\natexlab{c}})\citenamefont{Buend{\'\i}a, Rikvold, and
  Kolesik}}]{BUEN06B}
\bibinfo{author}{\bibfnamefont{G.~M.} \bibnamefont{Buend{\'\i}a}},
  \bibinfo{author}{\bibfnamefont{P.~A.} \bibnamefont{Rikvold}},
  \bibnamefont{and} \bibinfo{author}{\bibfnamefont{M.}~\bibnamefont{Kolesik}},
  \bibinfo{journal}{J.\ Mol.\ Struct.: THEOCHEM}
  \textbf{\bibinfo{volume}{769}}, \bibinfo{pages}{207}
  (\bibinfo{year}{2006}{\natexlab{c}}).

\bibitem[{\citenamefont{Marro and Dickman}(1999)}]{MARR99}
\bibinfo{author}{\bibfnamefont{J.}~\bibnamefont{Marro}} \bibnamefont{and}
  \bibinfo{author}{\bibfnamefont{R.}~\bibnamefont{Dickman}},
  \emph{\bibinfo{title}{Nonequilibrium Phase Transitions in Lattice Models}}
  (\bibinfo{publisher}{Cambridge University Press},
  \bibinfo{address}{Cambridge}, \bibinfo{year}{1999}).

%

\bibitem{KAMI93}
G.~Kamieniarz and H.~W.~J.\ Bl{\"o}te, J.\ Phys.\ A: Math.\ Gen.\ 
{\bf 26}, 201 (1993). 

\bibitem[{\citenamefont{Privman and Fisher}(1984)}]{PRIV84}
\bibinfo{author}{\bibfnamefont{V.}~\bibnamefont{Privman}} \bibnamefont{and}
  \bibinfo{author}{\bibfnamefont{M.~E.} \bibnamefont{Fisher}},
  \bibinfo{journal}{Phys.\ Rev.\ B} \textbf{\bibinfo{volume}{30}},
  \bibinfo{pages}{322} (\bibinfo{year}{1984}).

\bibitem[{\citenamefont{Privman}(1990)}]{PRIV90}
\bibinfo{author}{\bibfnamefont{V.}~\bibnamefont{Privman}}, in
  \emph{\bibinfo{booktitle}{Finite-Size Scaling and Numerical Simulation of
  Statistical Systems}}, edited by
  \bibinfo{editor}{\bibfnamefont{V.}~\bibnamefont{Privman}}
  (\bibinfo{publisher}{World Scientific}, \bibinfo{address}{Singapore},
  \bibinfo{year}{1990}).

\end{thebibliography}

\clearpage

%
%



\end{document}